\documentclass[aps,prl,preprint,tightenlines,showpacs,preprintnumbers,amsmath,amssymb,superscriptaddress]{revtex4}

\usepackage[dvips]{graphicx}
\begin{document}
\title{Finite frequency quantum noise in an interacting mesoscopic conductor}
\author{F.W.J. Hekking}
\affiliation{Laboratoire de Physique et Mod\'elisation des Milieux Condens\'es, CNRS
\& Universit\'e Joseph Fourier, BP 166, 38042 Grenoble-cedex 9, France}
\affiliation{Low Temperature Laboratory, Helsinki University of Technology, P.O. Box
3500, 02015 HUT, Finland}
\author{J.P. Pekola}
\affiliation{Low Temperature Laboratory, Helsinki University of Technology, P.O. Box
3500, 02015 HUT, Finland}


\begin{abstract}
We present a quantum calculation based on scattering theory of the frequency dependent
noise of current in an interacting chaotic cavity. We include interactions of the
electron system via long range Coulomb forces between the conductor and a gate with
capacitance $C$. We obtain explicit results exhibiting the two time scales of the
problem, the cavity's dwell time $\tau_D$ and the $RC$-time $\tau_C$ of the cavity
{\em vis \`a vis} the gate. The noise shows peculiarities at frequencies of the order
and exceeding the inverse charge relaxation time $\tau^{-1} = \tau^{-1}_D+\tau^{-1}_C
$.
\end{abstract}

\maketitle

Electrical noise in mesoscopic conductors is known to yield a wealth of information on
microscopic mechanisms of transport~\cite{blanter00a}. Recently, focus in theory and
experiment has shifted from studies of noise spectra to those of general current
statistics including all the moments of current~\cite{nazarov03}. Meanwhile, though,
important properties of the standard second moment, i.e., the part of statistics that
is normally associated with noise, are still to be uncovered in several important
aspects of direct relevance to experimental investigations. An example is finite
frequency noise of a mesoscopic conductor. The interest of studying it is two-fold.
First, one expects that the noise will probe the intrinsic dynamics of the conductor
at the Thouless energy (inverse dwell time $\tau_D$). Second, at finite frequency,
current is no longer spatially homogeneous, and charge piles up in the conductor.
Coulomb interaction screens this pile-up of charge, at a characteristic charge
relaxation frequency which may well be different from $1/\tau_D$. Therefore, any
theoretical treatment of finite-frequency noise should take interaction effects into
account~\cite{blanter00a}. Calculation of the noise spectrum of an arbitrary,
interacting mesoscopic conductor constitutes a formidable task, which has so far been
accomplished in specific cases only, like mesoscopic capacitors~\cite{blanter00a},
Coulomb blockade systems~\cite{coulblock} and Luttinger liquids~\cite{lutliq}. Finite
frequency noise spectra are of experimental relevance, as they can be measured with
recently developed quantum detectors~\cite{quantdet}. These noise detectors
distinguish finite positive and negative frequencies, i.e. they separately probe
emission and absorption.

\begin{figure}
    \includegraphics[width=8.3cm]{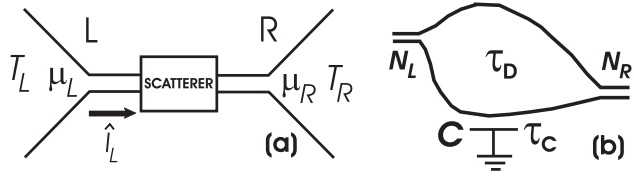}
    \caption{(a) Schematic presentation of the
two-terminal scattering problem. (b) Scheme of a chaotic quantum cavity; the capacitor
$C$ accounts for the interactions.}
    \label{scatteringcavity1}
\end{figure}

In this article we present a self-consistent theory of high frequency quantum noise of
a chaotic cavity. We calculate emission and absorption spectra with the help of
scattering theory, including interactions to ensure current conservation. Besides
providing timely guidelines for experiments on noise spectroscopy, our theory paves
the way towards a consistent calculation of the finite frequency higher moments of
current, in particular the presently actively investigated third
cumulant~\cite{thirdcum}. We obtain the noise spectrum of a chaotic quantum cavity
capacitively connected to a screening gate, see Fig.~\ref{scatteringcavity1}. We
believe these results to be representative, at least on a qualitative level, for an
arbitrary interacting conductor. Specifically we show that, under realistic
conditions, the noise spectrum is governed by the charge relaxation time $\tau$, where
$\tau^{-1}=\tau_D^{-1}+\tau_C^{-1}$, with the $RC$-time $\tau_C$ of the cavity. The
dwell time is $\tau_D=hN_F/N$, and $\tau_C=hC/(Ne^2)$, where $N_F$ is the density of
states of the cavity at the Fermi level, $N=N_L+N_R$ is the sum of the number of
channels of the left lead and right lead, respectively, and $C$ is the geometric
capacitance.

In the scattering theory of quantum transport the current operator is given
by~\cite{buttiker92}
\begin{eqnarray} \label{curalpha}
\hat I_\alpha (t)  =  \frac{e}{2\pi\hbar} \sum_n \int dE dE' e^{i(E-E')t/\hbar} \times
\nonumber
\\
\left[ \hat a^{\dagger}_{\alpha n} (E) \hat a_{\alpha n} (E') - \hat
b^{\dagger}_{\alpha n} (E) \hat b_{\alpha n} (E') \right].
\end{eqnarray}
The operator $\hat a^{\dagger}_{\alpha n} (E)$ ($\hat a$) creates (annihilates)
electrons incident upon the scatterer with energy $E$ in the transverse channel $n$ in
lead $\alpha$. Similarly, $\hat b^{\dagger}$, $\hat b$ denote electrons in the
outgoing states. For the two-terminal set-up depicted in
Fig.~\ref{scatteringcavity1}a, $\alpha$ takes values $L$ and $R$ for the left and
right leads respectively. The operators $\hat a$ and $\hat b$ are related by the
scattering matrix $s$, $\hat b_{\alpha n}(E) = s_{\alpha \gamma;nm}(E) \hat{a}_{\gamma
m}(E)$~\cite{footnote1}. The matrix $s$ is unitary and has dimensions $N \times N$.
Using $s$, we eliminate the $\hat{b}$ operators from Eq.~(\ref{curalpha}). Defining
the current matrix
\begin{eqnarray}
&& A^{(0)}_{\alpha \gamma ;mn}(L,E,E') = \nonumber \\
&& \delta_{mn} \delta_{L \alpha} \delta_{L \gamma} - s_{L \alpha;mk}^\dagger(E) s_{L
\gamma;kn}(E'), \label{curm}
\end{eqnarray}
we write current in the left lead as
\begin{eqnarray}
\label{leftcur} \hat I_L (t)  =  \frac{e}{2\pi\hbar} \int dE dE'
e^{i(E-E')t/\hbar} \times \nonumber \\
\hat a^{\dagger}_{\alpha m} (E) A^{(0)}_{\alpha \gamma ;mn}(L,E,E')\hat a_{\gamma n}
(E').
\end{eqnarray}
The superscript $(0)$ indicates that we deal with non-interacting electrons.

The average current-current correlation function in the left lead can now be written
as $\langle \delta \hat{I}_{L}(t) \delta\hat{I}_{L}(0)\rangle = \int \frac{d\omega}{2
\pi} e^{-i\omega t} S_{LL}(\omega)$, where we denote $\delta \hat{I}_{L}(t) \equiv
\hat{I}_{L}(t) - \langle \hat{I}_{L}(t)\rangle$. Using Eqs.~(\ref{curm}) and
(\ref{leftcur}), and performing standard thermal averaging of products of operators
$\hat{a}$ and $\hat{a}^\dagger$ for equilibrium reservoirs, we obtain the noise
spectrum in the absence of interactions~\cite{blanter00a}
\begin{eqnarray}
S^{(0)}_{LL}(\omega) = \frac{e^2}{h} \int dE \mbox{Tr}[A^{(0)}_{\alpha
\gamma}(L,E,E+\hbar \omega) A^{(0)}_{\gamma \alpha}(L,E+\hbar \omega,E)] \times \nonumber\\
\left\{f_\alpha(E)[1-f_\gamma(E+\hbar\omega)] + [1-f_\alpha(E)]f_\gamma(E+\hbar
\omega)\right\}. \label{dIdI}
\end{eqnarray}
Here $f_\alpha (E)$ is the Fermi function in lead $\alpha$; the notation
$A^{(0)}_{\alpha \gamma}(L,E,E')$ indicates the $N_\alpha \times N_\gamma$ block of
the matrix $A^{(0)}(L,E,E')$; $\mbox{Tr} [C_{\alpha \alpha}]$ is the trace of the
square matrix $C_{\alpha \alpha}$ which implies summing over the $N_\alpha$
propagating channels of lead $\alpha$. For energy-independent scattering, the
integration over $E$ can be straightforwardly performed and Eq.~(\ref{dIdI}) reduces
to the result found in~\cite{khlus87}, valid for a tunnel barrier or a quantum point
contact. However, quite generally the current matrix $A$ contains correlations at
different energies separated by $\hbar \omega$~\cite{reulet05}. Here we consider the
effect of these correlations on the finite-frequency noise of a chaotic quantum
cavity, which can be treated in the framework of random matrix theory (RMT).
Specifically, we make use of~\cite{polianski03} to obtain the ensemble averaged,
energy-dependent quantity $\langle \mbox{Tr}[A^{(0)}_{\alpha \gamma}(L,E,E+\hbar
\omega) A^{(0)}_{\gamma \alpha}(L,E+\hbar \omega,E)] \rangle$ (no sum over $\alpha$
and $\gamma$). In view of Eq.~(\ref{curm}), this amounts to the calculation of the
average trace of the product of two scattering matrices, which we achieve with the
help of~\cite{polianski03}, $\langle \mbox{Tr}[s_{\alpha \gamma}^\dagger (E_1)
s_{\alpha \gamma} (E_2)] \rangle \equiv (N_\alpha N_\gamma/N) g(E_2 -E_1)$, where $g
(E) = [1-iE\tau_D/\hbar]^{-1}$. Similarly, we need
\begin{eqnarray}
 \langle \mbox{Tr}[s_{\gamma \alpha}^\dagger (E_1) s_{\gamma \delta} (E_2) s _{\varepsilon
\delta }^\dagger(E_3) s_{\varepsilon \alpha} (E_4)] \rangle = \frac{N_\alpha
N_\gamma}{N^2} \left[\vphantom{\int }\delta _{\alpha \delta} N_\varepsilon g(E_2-E_1)
g (E_4 - E_3) + \right.
\nonumber \\
\left. \delta _{\gamma \varepsilon} N_\delta g(E_4 - E_1) g(E_2 - E_3) -
\frac{N_\delta N_\varepsilon}{N}\frac{g(E_2-E_1) g (E_4 - E_3) g(E_4 - E_1) g(E_2 -
E_3)}{g(E_2 +E_4 - E_1 - E_3)}\right] . \nonumber
\end{eqnarray}
These results depend on the cavity's dwell-time $\tau_D$ as well as on the number of
channels $N_{L,R}$ in the leads. Certain inequalities apply to these quantities. First
of all, we assume $\tau_D$ to be much longer than the Ehrenfest time that
characterizes the spreading of a wave packet in the cavity~\cite{agam00}. This ensures
that the electrons spend enough time in the cavity for quantum chaos to fully develop
and RMT to apply. Hence our results are universal in the sense that they do not depend
on details of the cavity. Second, the results of~\cite{polianski03} were obtained for
a large scattering matrix, i.e., in the limit $N \gg1$. We ignore corrections of
relative order ${\cal O}(1/N)$, i.e., weak localization is not taken into account.
Finally, we also ignore renormalization of transport by interactions~\cite{reneff},
assuming that both leads are well-connected to the cavity with $N_L, N_R \gg 1$.
However, the openings to the leads are much smaller than the total circumference of
the cavity in order for universality to hold.

Performing the ensemble-average of~(\ref{dIdI}), we find the frequency-dependent noise
of a non-interacting chaotic cavity at finite temperature $T$ and applied bias $V$,
\begin{eqnarray}
S^{(0)}_{LL}(\omega) = &&\frac{e^2}{h} \frac{1}{1+\omega^2 \tau_D^2}
\left\{\frac{2\hbar \omega}{1 - e^{-\beta \hbar \omega}} \left[\frac{N_L
N_R}{N}\left(1-\frac{N_L N_R}{N^2}\right) + \omega ^2 \tau_D^2\frac{N_L}{2} \left(1
+\frac{N_L^2}{N^2} +  \frac{N_R^2}{N^2}\right)\right] + \right. \nonumber\\
&&\left. \left[\frac{\hbar \omega + eV}{1 - e^{-\beta (\hbar \omega + eV)}}  +
\frac{\hbar \omega - eV}{1 - e^{-\beta (\hbar \omega - eV)}}\right] \left[\frac{N_L^2
N_R^2}{N^3} + \omega^2 \tau_D^2\frac{N_L^2N_R}{N^2} \right] \right\}, \label{cavcum0}
\end{eqnarray}
where $\beta = 1/k_B T$. In the limit $\omega \to 0$, Eq.~(\ref{cavcum0}) reduces to
the known result for zero-frequency noise of a chaotic cavity~\cite{zerofreqcav} which
has been studied experimentally~\cite{oberholzer01}. We will discuss the
frequency-dependent spectral function below, after interaction effects have been
included.

In order to treat the case of an interacting chaotic cavity we follow the general
framework of Ref.~\cite{pedersen98}. The central idea is that fluctuations of particle
current, as given by Eq.~(\ref{curalpha}), lead to a pile-up of charge in the cavity
that is subsequently compensated by displacement currents induced at the contacts and
the gate. The total current at a given contact can be written as the sum of particle
current and displacement current at that contact. To determine the displacement
current, the uniform self-consistent potential of the cavity must be
found~\cite{selfcon}, a problem that can be solved
explicitly~\cite{pedersen98,constpot}. It turns out that the resulting total current
operator for the left lead is still given by Eq.~(\ref{leftcur}), however the current
matrix $A^{(0)}$ should be replaced by an effective current matrix $A = A^{(0)} +
\Delta A$, where the correction $\Delta A$ includes the displacement contribution.
This approach guarantees current conservation at finite frequency: the sum of all
currents at the contacts of the sample together with the displacement current at the
gate is zero. The explicit form of the correction $\Delta A$ reads~\cite{pedersen98}
\begin{equation}
    \Delta A_{\gamma \delta}(\alpha,E,E+\hbar\omega) =
  i 2 \pi \hbar \omega {\cal G}_\alpha (\omega)
   {\cal N}_{\gamma \delta}(E,E+\hbar \omega),
\end{equation}
where
\begin{equation}
{\cal N}_{\gamma \delta}(E,E') = \frac{i}{2\pi} \frac{s_{\alpha\gamma}^\dagger(E)
\left[ s_{\alpha\delta}(E)- s_{\alpha\delta}(E')\right] }{E'-E} \label{density_matrix}
\end{equation}
is the density of states matrix and ${\cal G}_\alpha (\omega)$ a frequency dependent
response function, proportional to the emittance~\cite{buttiker93} of the cavity. In
the linear response regime,
\begin{equation}
{\cal G}_\alpha (\omega) = \frac{N_\alpha}{N}\frac{e^2 N_F}{C + e^{2} N_F} \frac{1}{1
- i\omega \tau} ,
\end{equation}
where the frequency-dependence is governed by the charge relaxation time $\tau$.

Turning to the noise, we replace $A^{(0)}$ by $A$ in Eq.~(\ref{dIdI}) and find the
additional contributions to noise related to the interaction correction $\Delta A$.
Performing the ensemble averaging of these terms ($\langle A^{(0)} \Delta A \rangle$,
$\langle \Delta A A^{(0)} \rangle $ and $\langle \Delta A \Delta A\rangle$), we obtain
the full noise spectral function: {\em it is given by Eq.~(\ref{cavcum0}), upon
replacing the dwell time $\tau_D$ by the charge relaxation time $\tau$}. This is our
central result, which we will discuss in several relevant limits.

\begin{figure}
    \includegraphics[width=8.3cm]{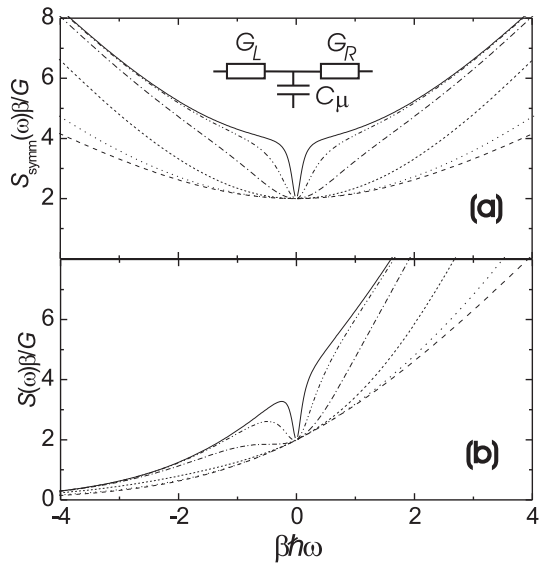}
    \caption{(a) Frequency-dependent symmetrized noise $S_{symm}(\omega)$ in units $G/\beta$ of an
    interacting chaotic cavity ($N_L/N=0.5$) with a few values of charge
relaxation
    time: from top $\tau/(\hbar\beta)=10,3,1,0.3,0.1$; the bottom one corresponds to
    energy-independent scattering $\tau =0$.
    Inset: corresponding circuit model. (b) Frequency-dependent
    unsymmetrized quantum noise; parameters as in (a).}
    \label{Fig2}
\end{figure}

If $V\to 0$, we find the non-symmetrized, equilibrium noise spectrum,
\begin{equation}
S_{LL}(\omega) = \frac{e^2}{h} \frac{N_L N_R}{N}
 \frac{2\hbar \omega}{1 - e^{-\beta \hbar \omega}}
\frac{1 + \omega ^2 \tau^2 N/N_R}{1+\omega^2 \tau^2}. \label{nonsymS}
\end{equation}
The symmetrized spectrum $[S_{LL}(\omega)+S_{LL}(-\omega)]/2$ is plotted in
Fig.~\ref{Fig2}a. According to the fluctuation dissipation theorem, this is connected
to the real part of the AC conductance of the interacting cavity. We find
$$
\Re\mbox{e}[G_{LL}(\omega)] = \frac{e^2}{h}\frac{N_L N_R}{N} \frac{1 +  \omega ^2
\tau^2 N/N_R}{1 + \omega^2 \tau^2},
$$
in agreement with a direct calculation~\cite{brouwer97}. At equilibrium, the cavity
corresponds to the effective circuit depicted in the inset of Fig.~\ref{Fig2}a. Here,
$G_L$ and $G_R$ are the quantum conductances of the left and right lead respectively,
whereas $C_\mu$ is the series capacitance of the true geometric capacitance $C$ and
the quantum capacitance $e^2 N_F$ of the cavity. At low frequencies, the effect of
this capacitance can be ignored, and the noise is that of the series conductance $G =
G_LG_R/(G_L +G_R) = (N_L N_R/N)e^2/h$. At high frequency, $C_\mu$ opens a new current
path thereby increasing the AC conductance and hence the noise. The corresponding
frequency scale is set by $\tau^{-1}$, which is the inverse $RC$ time of the two
resistances $G_L^{-1}$ and $G_R^{-1}$ in parallel, in combination with the capacitance
$C_\mu$. Quite generally, the non-interacting case corresponds to the limit of large
geometric capacitance, such that the quantum capacitance -- and hence the dwell time
$\tau_D$-- determines the dynamics. In this limit the noise spectra are obtained from
Eq.~(\ref{cavcum0}). The result~(\ref{nonsymS}) for non-symmetrized noise is plotted
in Fig.~\ref{Fig2}b. There is a pronounced difference between noise at negative and
positive frequencies, corresponding respectively to absorption and emission. For
negative frequencies a peculiarity develops at $\omega \sim -\tau^{-1}$ as a
manifestation of enhanced absorption due to charged fluctuations.

\begin{figure}
    \includegraphics[width=8.3cm]{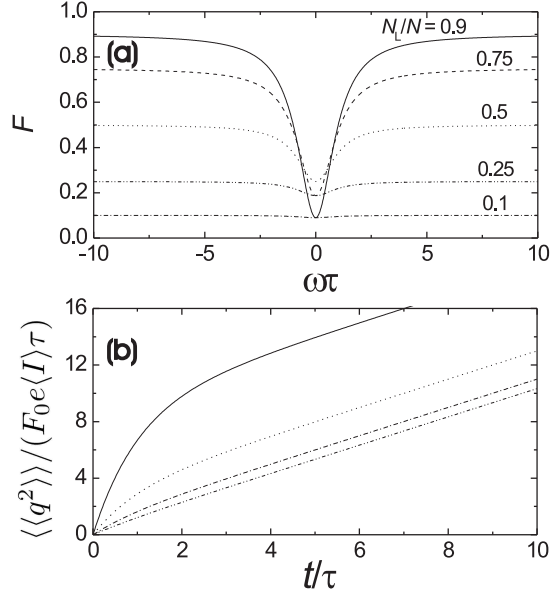}
    \caption{(a) Frequency dependence of
    the Fano factor at a few values of $N_L/N$.
    (b) The second cumulant for a cavity with
    $N_L/N=0.9, 0.75, 0.5,$ and $0.3$ from top to bottom.}
    \label{Fig3}
\end{figure}
As to the non-equilibrium noise for $eV \gg k_B T,\hbar \omega$ we have
\begin{equation} \label{shotint}
S_{LL}(\omega) = \frac{N_L N_R}{N^2} e\langle I \rangle
 \frac{1 + \omega ^2 \tau^2 N/N_R}{1+\omega^2 \tau^2},
\end{equation}
where we defined the average current through the cavity $\langle I \rangle = G V$. The
Fano factor $F(\omega) \equiv S_{LL}(\omega)/(e\langle I \rangle)$ is plotted in
Fig.~\ref{Fig3}a. At zero frequency this gives the result $S_{LL}(0) = F(0) e\langle I
\rangle$, which is characterized by the shot-noise reduction factor $F(0)=N_L
N_R/N^2$, in agreement with~\cite{zerofreqcav}. At frequencies exceeding $1/\tau$, the
Fano factor tends to the value $N_L/N$, corresponding to an increase of the noise by
$N/N_R$.

We finally consider the time-dependent second cumulant of charge
transport~\cite{levitov96} for the left lead, given by
\begin{equation} \label{secondcum}
\langle \langle q^2 \rangle \rangle(t) = 2 \int \frac{d\omega}{2\pi} S_{LL}(\omega)
\frac{1-\cos (\omega t)}{\omega^2}.
\end{equation}
Inserting (\ref{shotint}) into (\ref{secondcum}) we have
\begin{equation}
\langle \langle q^2 \rangle \rangle (t)= \frac{N_L N_R}{N^2} e\langle I \rangle
t\left[1+ \frac{N_L}{N_R}\frac{\tau}{t}(1-e^{-t/\tau})\right].
\end{equation}
The time dependence of $\langle \langle q^2\rangle \rangle/(F_0e\langle I
\rangle\tau)$ is plotted in Fig.~\ref{Fig3}b again at a few values of $N_L/N$. We see
that the universal long time limit is reached only after a transient time $\tau$; the
ratio of the slopes at early and late times is determined by the asymmetry and given
by $N/N_R$.

Finally, we comment on the experimental feasibility of a measurement of the above
found noise spectra. For the cavity studied in the experiment~\cite{oberholzer01}, the
various conditions for our calculations to be valid appear to be met. The dwell time
$\tau _D$ is in the ns range;  the large area, $> 10$ ($\mu$m)$^2$ for a typical
cavity, allows for capacitances exceeding 10 fF. We thus estimate $\tau _C \sim 20$
ps, thereby giving a charge relaxation time $\tau$ of the same order. The condition
$\omega \tau \sim 1$ can then be reached, using a Josephson junction in the quantum
limit as a detector, where the level spacing (plasma frequency) is $ \sim$ 10...100
GHz. Our results may be helpful for future experiments on other mesoscopic conductors
also, such as a diffusive metallic wire capacitively coupled to a nearby ground
plane~\cite{footnote2}.

We thank T. Heikkil\"a, F. Pistolesi, R. Whitney and especially P. Brouwer, M.
B\"uttiker and D. Prober for useful discussions and comments. Financial support from
Academy of Finland and from Institut Universitaire de France is gratefully
acknowledged.


\begin{thebibliography}{99}
\bibitem{footnote1} We sum over repeated indices unless stated
otherwise.
\bibitem{footnote2} Here, a more elaborate treatment of screening may be needed as the uniform potential model breaks down~\cite{constpot}.
\end{thebibliography}
\end{document}